\documentclass[journal]{IEEEtran}

\ifCLASSINFOpdf
\else
\fi

\usepackage[colorlinks=true]{hyperref}
\usepackage{hyperref}
\usepackage{graphicx}
\usepackage{amsmath,amssymb} % define this before the line numbering.
\usepackage{color}
\usepackage{xcolor}
\usepackage[font=small,labelfont=bf]{caption}

\usepackage{makeidx}  % allows for indexgeneration
\usepackage{color}
\usepackage{graphicx}

\usepackage{amsmath}
\usepackage{amssymb}
\usepackage[labelformat=simple]{subfig}
\usepackage[ruled,lined,linesnumbered]{algorithm2e}
\usepackage{algorithmic}
\usepackage{blindtext}
\usepackage{enumitem}
\usepackage{multirow}
\usepackage[labelformat=simple]{subfig}
\usepackage{threeparttable/threeparttable}
\usepackage{bm}
\usepackage[normalem]{ulem}
\usepackage{hyperref}
\usepackage[utf8]{inputenc}
\usepackage{bibunits}

\usepackage{colortbl}
\definecolor{maroon}{cmyk}{0,0.87,0.68,0.32}
\definecolor{brown(traditional)}{rgb}{0.59, 0.29, 0.0}

\usepackage{amssymb}% http://ctan.org/pkg/amssymb
\usepackage{pifont}% http://ctan.org/pkg/pifont
\newcommand{\cmark}{\ding{51}}%
\newcommand{\xmark}{\ding{55}}%

\newcommand{\etal}{\mbox{\emph{et al.}}}
\newcommand{\ie}{\mbox{\emph{i.e.,\ }}}

\newcommand{\bd}[1]{\textbf{#1}}

\usepackage{array}
\newcolumntype{P}[1]{>{\centering\arraybackslash}p{#1}}

\newenvironment{jlrevrev}{}{}
\newenvironment{jlrev}{}{}

\newenvironment{jlblue}{\color{black}}{}

\definecolor{zlue}{rgb}{0.22, 0.09, 1}

\begin{document}

\title{UNet++: Redesigning Skip Connections to Exploit Multiscale Features in Image Segmentation}

\author{Zongwei~Zhou,~\IEEEmembership{Member,~IEEE,}
        Md~Mahfuzur~Rahman~Siddiquee,~\IEEEmembership{Member,~IEEE,} \\
        Nima~Tajbakhsh,~\IEEEmembership{Member,~IEEE,}
        and Jianming~Liang,~\IEEEmembership{Senior Member,~IEEE}% <-this % stops a space
\thanks{Z. Zhou, N. Tajbakhsh and J. Liang are with the Department of Biomedical Informatics, Arizona State University, Scottsdale, AZ 85259 USA.\newline (\href{mailto:zongweiz@asu.edu}{zongweiz@asu.edu};~\href{mailto:ntajbakh@asu.edu}{ntajbakh@asu.edu};~\href{mailto:jianming.liang@asu.edu}{jianming.liang@asu.edu})}% <-this % stops a space
\thanks{M M. Rahman Siddiquee is with School of Computing, Informatics, and Decision Systems Engineering, Arizona State University, Tempe, AZ 85281 USA. (\href{mailto:mrahmans@asu.edu}{mrahmans@asu.edu})}
\thanks{\noindent Published in IEEE Transactions on Medical Imaging}}
%\thanks{Copyright (c) 2019 IEEE. Personal use of this material is permitted. However, permission to use this material for any other purposes must be obtained from the IEEE by sending a request to \url{pubs-permissions@ieee.org}.}}
%\thanks{Manuscript received xxx; revised xxx.}}

\markboth{Journal of IEEE Transactions on Medical Imaging}%
{Zhou \MakeLowercase{\textit{et al.}}: Deeply Supervised, Nested Ensemble of Convolutional Neural Nets for Medical Image Segmentation}



\maketitle


% The motivation of the paper is not semantic gap anymore, but about multiple level of features aggregation. In doing so, we ensemble different depths of U-Net together and redesign the skip connection.
\begin{abstract}
The state-of-the-art models for medical image segmentation are variants of U-Net and fully convolutional networks (FCN). Despite their success, these models have two  limitations: (1) their optimal depth is apriori unknown, requiring extensive architecture search or inefficient ensemble of models of varying depths; and (2) their skip connections impose an unnecessarily restrictive fusion scheme, forcing aggregation only at the same-scale feature maps of the encoder and decoder sub-networks. To overcome these two limitations, we propose UNet++, a new neural architecture for semantic and instance segmentation, by (1) alleviating the unknown network depth with an efficient ensemble of U-Nets of varying depths, which partially share an encoder and co-learn simultaneously using deep supervision; (2) redesigning skip connections to  aggregate features of varying semantic scales at the decoder sub-networks, leading to a highly flexible feature fusion scheme; and (3) devising a pruning scheme to accelerate the inference speed of UNet++. We have evaluated UNet++ using six different medical image segmentation datasets, covering multiple imaging modalities such as computed tomography (CT), magnetic resonance imaging (MRI), and electron microscopy (EM), and demonstrating that (1) UNet++ consistently outperforms the baseline models for the task of semantic segmentation across different datasets and backbone architectures; (2) UNet++ enhances segmentation quality of varying-size objects---an improvement over the fixed-depth U-Net; \iffalse that is bound to segment objects of only certain sizes;\fi (3) Mask RCNN++ (Mask R-CNN with UNet++ design) outperforms the original Mask R-CNN for the task of instance segmentation; and (4) pruned UNet++ models achieve significant speedup while showing only modest performance degradation.  Our implementation and pre-trained models are available at
\href{https://github.com/MrGiovanni/UNetPlusPlus}{https://github.com/MrGiovanni/UNetPlusPlus}.
\end{abstract}


% \begin{bibunit}[plain]

% \fi
\IEEEpeerreviewmaketitle

\section{Introduction}
\label{sec:introduction}

The encoder-decoder networks are widely used in modern semantic and instance segmentation models~\cite{zhou2017deep,shen2017deep,litjens2017survey,chartrand2017deep,falk2018u,tajbakhsh2019embracing}. Their success is largely attributed to their skip connections, which combine deep, semantic, coarse-grained feature maps from the decoder sub-network with shallow, low-level, fine-grained feature maps from the encoder sub-network, and have proven to be effective in recovering fine-grained details of the target objects~\cite{drozdzal2016importance,he2016deep,huang2017densely} even on complex background~\cite{hariharan2015hypercolumns,lin2017feature}. Skip connections have also played a key role in the success of instance-level segmentation models such as~\cite{he2017mask,hu2018learning} where the idea is to segment and distinguish each instance of desired objects. 

However, these encoder-decoder architectures for image segmentation come with two limitations. First, the optimal depth of an encoder-decoder network can vary from one application to another, depending on the task difficulty and the amount of labeled data available for training. A simple approach would be to train models of varying depths separately and then ensemble the resulting models during the inference time~\cite{dietterich2000ensemble,hoo2016deep,ciompi2015automatic}. However, this simple approach is inefficient from a deployment perspective, because these networks do not share a common encoder. Furthermore, being trained independently, these networks do not enjoy the benefits of multi-task learning~\cite{bengio2009learning,zhang2017survey}.
Second, the design of skip connections used in an encoder-decoder network is unnecessarily restrictive, demanding the fusion of the same-scale encoder and decoder feature maps. While striking as a natural design,  the same-scale feature maps from the decoder and encoder networks are semantically dissimilar and no solid theory guarantees that they are the best match for feature fusion. 

%##############################################################################################
\begin{figure*}[t]
\begin{center}
\includegraphics[width=1.0\linewidth]{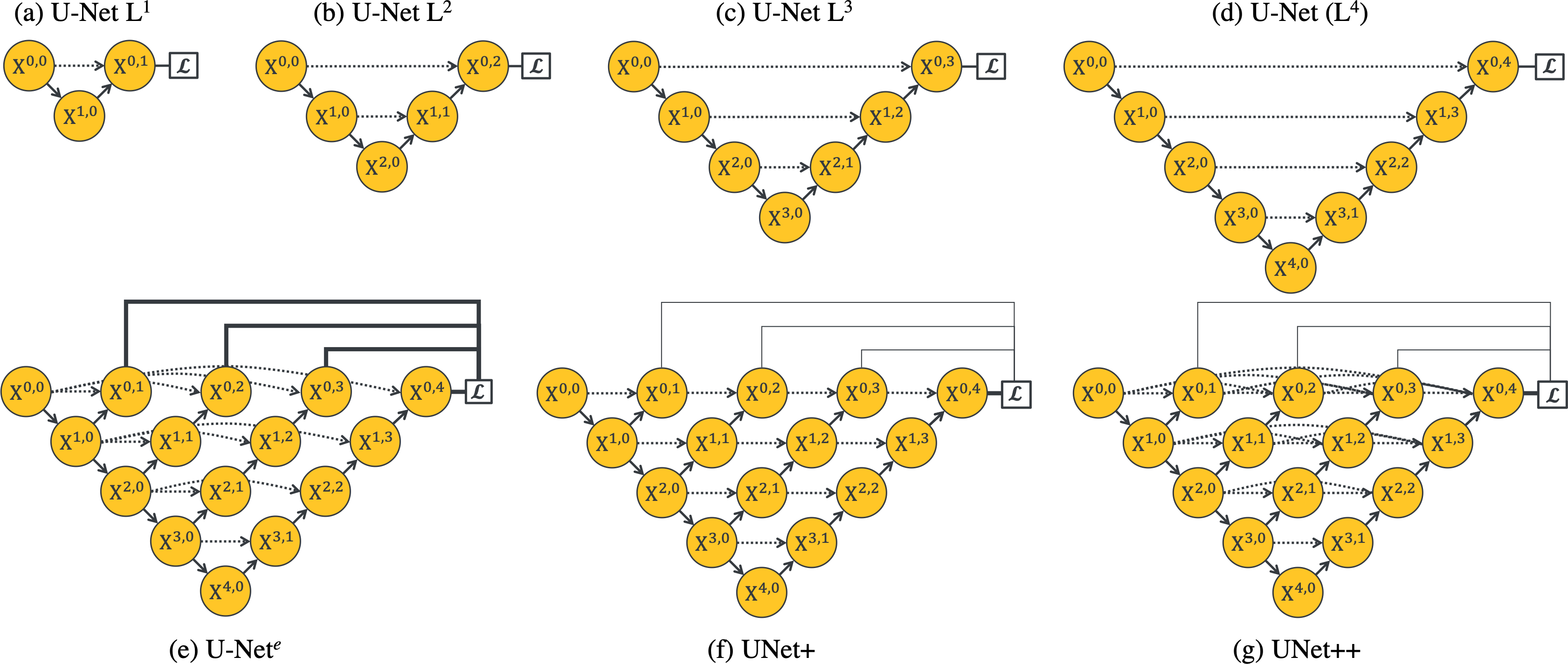}
\end{center}
\caption{Evolution from U-Net to UNet++. Each node in the graph represents a convolution block, downward arrows indicate down-sampling, upward arrows indicate up-sampling, and dot arrows indicate skip connections. (a--d) U-Nets of varying depths. (e) Ensemble architecture, U-Net$^e$, which combines U-Nets of varying depths into one unified architecture. All U-Nets (partially) share the same encoder, but have their own decoders. (f) UNet+ is constructed from U-Net$^e$ by dropping the original skip connections and connecting every two adjacent nodes with a short skip connection, enabling the deeper decoders to send supervision signals to the shallower decoders. (g) UNet++ is constructed from U-Net$^e$ by connecting the decoders, resulting in densely connected skip connections, enabling dense feature propagation along skip connections and thus more flexible feature fusion at the decoder nodes. As a result, each node in the UNet++ decoders, from a horizontal perspective, combines multiscale features from its all preceding nodes at the same resolution, and from a vertical perspective, integrates multiscale features across different resolutions from its preceding node, as formulated at Eq.~\ref{eq_unet}. This multiscale feature aggregation of UNet++ gradually synthesizes the segmentation, leading to increased accuracy and faster convergence, as evidenced by our empirical results in~\sectionname~\ref{sec:results}.
Note that, explicit deep supervision is required (bold links) to train U-Net$^e$ but optional (pale links) for UNet+ and UNet++.}
\label{fig:network_architecture}
\end{figure*}
%##############################################################################################

In this paper, we present UNet++, a new general purpose image segmentation architecture that aims at overcoming the above limitations. {\jlrevrev As presented in~\figurename~\ref{fig:network_architecture}(g), UNet++ consists of U-Nets of varying depths whose decoders are densely connected at the same resolution via the redesigned skip connections. The architectural changes introduced in UNet++ enable the following advantages.} First, UNet++ is not prone to the choice of network depth because it embeds U-Nets of varying depths in its architecture. All these U-Nets partially share an encoder, while their decoders are intertwined. By training UNet++ with deep supervision, all the constituent U-Nets are trained simultaneously while benefiting from a shared image representation. This design not only improves the overall segmentation performance, but also enables model pruning during the inference time. Second, UNet++ is not handicapped by unnecessarily restrictive skip connections where only the same-scale feature maps from the encoder and decoder can be fused. The redesigned skip connections introduced in  UNet++ present feature maps of varying scales at a decoder node, allowing the aggregation layer to decide how various feature maps carried along the skip connections should be fused with the decoder feature maps. The redesigned skip connections are realized in UNet++ by densely connecting the decoders of the constituents U-Nets at the same resolution.
We have extensively evaluated UNet++ across six segmentation datasets and multiple backbones of different depths. Our results demonstrate that UNet++ powered by redesigned skip connections and deep supervision enables a significantly higher level of performance for both semantic and instance segmentation. 
This significant improvement of UNet++ over the classical U-Net architecture is ascribed to the advantages offered by the redesigned skip connections and the extended decoders, which together enable gradual aggregation of the image features  across the network, both horizontally and vertically.

In summary, we make the following five contributions:
\begin{enumerate}[noitemsep]
  \item We introduce a built-in ensemble of U-Nets of varying depths in UNet++, enabling improved segmentation performance for varying size objects---an improvement over the fixed-depth U-Net \iffalse that is bound to segment objects of certain sizes\fi (see \sectionname~\ref{sec:model}).
  \item We redesign skip connections in UNet++, enabling flexible feature fusion in decoders---an improvement over the restrictive skip connections in U-Net that require fusion of only same-scale feature maps (see \sectionname~\ref{sec:model}).
  \item We devise a scheme to prune a trained UNet++, accelerating its inference speed while maintaining its performance (see \sectionname~\ref{sec:model_pruning}).
  \item We discover that simultaneously training multi-depth U-Nets embedded within the UNet++ architecture stimulates collaborative learning among the constituent U-Nets, leading to much better performance than individually training isolated U-Nets of the same architecture (see \sectionname~\ref{sec:embedded_vs_isolated_training} and \sectionname~\ref{sec:collarborative_learning}).
  \item We demonstrate the extensibility of UNet++ to multiple backbone encoders and further its applicability to various medical imaging modalities including CT, MRI, and electron microscopy (see \sectionname~\ref{sec:semantic_results} and \sectionname~\ref{sec:instance_segmentation}).
\end{enumerate}

%##############################################################################################
\begin{figure*}[t]
\begin{center}
\includegraphics[width=1.0\linewidth]{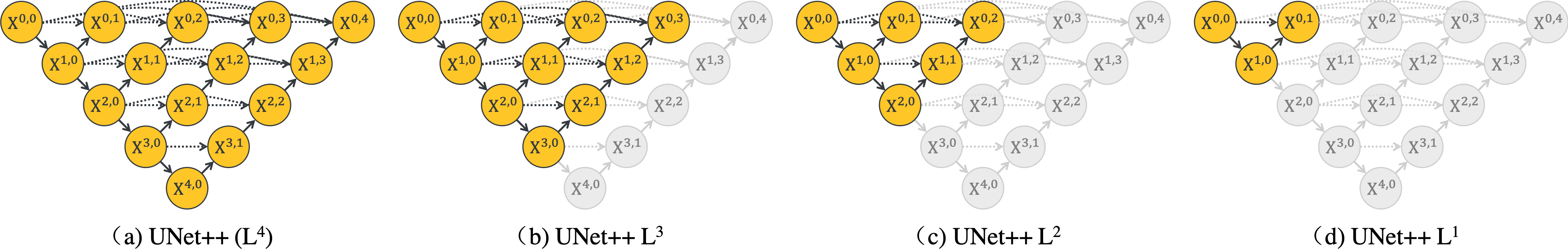}
\end{center}
\caption{Training UNet++ with deep supervision makes segmentation results available at multiple nodes X$^{0,j}$, enabling architecture pruning at inference time. Taking the segmentation result from X$^{0,4}$ leads to no pruning, UNet++ (L$^4$),  whereas taking the segmentation result from X$^{0,1}$ results in a maximally pruned architecture, UNet++ L$^1$. Note that nodes removed during pruning are colored in gray. \iffalse For further details, see  \sectionname~\ref{sec:model_pruning}.  The pruned components (in gray) are independent with the remaining parts (in yellow) at inference time, but contribute during training.\fi}
\label{fig:prune_structure}
\end{figure*}
%##############################################################################################

\section{Proposed Network Architecture: UNet++}
\label{sec:methods}
\figurename~\ref{fig:network_architecture} shows how UNet++ evolves from the original U-Net. In the following, we first trace this evolution, motivating the need for UNet++, and then explain its technical and implementation details.

\subsection{Motivation behind the new architecture}
\label{sec:motiv}

We have done a comprehensive ablation study to investigate the performance of U-Nets of varying depths (\figurename~\ref{fig:network_architecture}(a-d)). For this purpose, we have used three relatively small datasets, namely \texttt{Cell}, \texttt{EM}, and \texttt{Brain Tumor} (detailed in~\sectionname~\ref{sec:dataset}). \tableautorefname~\ref{tab:architecture_analysis} summarizes the results. For the cell and brain tumor segmentation, a shallower network (U-Net L$^3$) outperforms the deep U-Net. For the EM dataset, on the other hand, the deeper U-Nets consistently outperform the shallower counterparts, but the  performance gain is only marginal. Our experimental results suggest two key findings: 1) deeper U-Nets are not necessarily always better, 2) the optimal depth of architecture depends on the difficulty and size of the dataset at hand. While these findings may encourage an automated neural architecture search, such an approach is hindered by the limited computational resources~\cite{liu2018progressive,zoph2018learning,liu2019auto,zhang2019customizable,li2019partial}. Alternatively, we propose an ensemble architecture, which combines  U-Nets of varying depths into one unified structure. We refer to this architecture as U-Net$^e$ (\figurename~\ref{fig:network_architecture}(e)). 
We train U-Net$^e$ by defining a separate loss function for each U-Net in the ensemble, \ie X$^{0,j},~j\in\{1,2,3,4\}$. Our deep supervision scheme differs from the commonly used deep supervision in deep image classification and image segmentation networks; in ~\cite{xie2015holistically,chen2016deep,dou20173d,lee2015deeply} the auxiliary loss functions are added to the nodes along the decoder network, \ie X$^{4-j,j},~j\in\{0,1,2,3,4\}$, whereas we apply them on X$^{0,j},~j\in\{1,2,3,4\}$. At the inference time, the output from each U-Net in the ensemble is averaged.

%##############################################################################################
\begin{table}[t]
\footnotesize
\begin{center}
\begin{threeparttable}
\caption{{\jlrev Ablation study on U-Nets of varying depths alongside with the new variants of U-Nets proposed in this work. U-Net L$^d$ refers to a U-Net with a depth of $d$ (\figurename~\ref{fig:network_architecture}(a-d)). U-Net$^e$, UNet+, and UNet++ are the new variants of U-Net, which are depicted in \figurename~\ref{fig:network_architecture}(e-g). ``DS'' denotes deeply supervised training followed by average voting. Intersection over union (IoU) is used as the metric for comparison (mean$\pm$s.d. \%).}}
\label{tab:architecture_analysis}
    {\jlrev
    \begin{tabular}{p{0.15\linewidth}P{0.04\linewidth}P{0.11\linewidth}P{0.12\linewidth}P{0.1\linewidth}P{0.19\linewidth}}
    \hline
    \textbf{Architecture} & \textbf{DS} & \textbf{Params} & \textbf{EM} & \textbf{Cell} & \textbf{Brain Tumor} \\
    \hline
    U-Net L$^1$  & \xmark & 0.1M & 86.83{\tiny $\pm$0.43} & 88.58{\tiny $\pm$1.68} & 86.90{\tiny $\pm$2.25}  \\
    U-Net L$^2$  & \xmark & 0.5M & 87.59{\tiny $\pm$0.34} & 89.39{\tiny $\pm$1.64} &  88.71{\tiny $\pm$1.45} \\
    U-Net L$^3$   & \xmark & 1.9M & 88.16{\tiny $\pm$0.29} & 90.14{\tiny $\pm$1.57} & 89.62{\tiny $\pm$1.41}  \\
    U-Net (L$^4$)  & \xmark & 7.8M & 88.30{\tiny $\pm$0.24} & 88.73{\tiny $\pm$1.64} & 89.21{\tiny $\pm$1.55} \\
    U-Net$^e$  & \cmark  & 8.7M & 88.33{\tiny $\pm$0.23} & 90.72{\tiny $\pm$1.51} &  90.19{\tiny $\pm$0.83} \\
    \hline
    UNet+  & \xmark & 8.7M & 88.39{\tiny $\pm$0.15} & 90.71{\tiny $\pm$1.25} & 90.70{\tiny $\pm$0.91}  \\
    UNet+  & \cmark & 8.7M & 88.89{\tiny $\pm$0.12} & 91.18{\tiny $\pm$1.13} & 91.15{\tiny $\pm$0.65} \\
    \hline
    UNet++  & \xmark & 9.0M & 88.92{\tiny $\pm$0.14} & 91.03{\tiny $\pm$1.34} & 90.86{\tiny $\pm$0.81}\\
    UNet++  & \cmark & 9.0M & \textbf{89.33{\tiny $\pm$0.10}} & \textbf{91.21{\tiny $\pm$0.98}} & {\bf 91.21{\tiny $\pm$0.68}} \\
    \hline
    \end{tabular}
    }
\end{threeparttable}
\end{center}
\end{table}
%##############################################################################################

The ensemble architecture (U-Net$^e$) outlined above benefits from knowledge sharing, because all U-Nets within the ensemble partially share the same encoder even though they have their own  decoders. However, this architecture still suffers from two drawbacks. First, the decoders are disconnected---deeper U-Nets do not offer a supervision signal to the decoders of the shallower U-Nets in the ensemble. Second, the common design of skip connections used in the U-Net$^e$ is unnecessarily restrictive, requiring the network to combine the decoder feature maps with only the same-scale feature maps from the encoder. While striking as a natural design,  there is no guarantee that the same-scale feature maps are the best match for the feature fusion. 

To overcome the above limitations, we remove original skip connections from the U-Net$^e$ and connect every two adjacent nodes in the ensemble, resulting in a new architecture, which we refer to as UNet+ (\figurename~\ref{fig:network_architecture}(f)). Owing to the new connectivity scheme, UNet+ connects the disjoint decoders, enabling gradient back-propagation from the deeper decoders to the shallower counterparts. UNet+ further relaxes the unnecessarily restrictive behaviour of skip connections by presenting each node in the decoders with the aggregation of all feature maps computed in the shallower stream. While using aggregated feature maps at a decoder node is far less restrictive than having only the same-scale feature map from the encoder,  there is still room for improvement. We further propose to use dense connectivity in UNet+,  resulting in our final architecture proposal, which we refer to as UNet++ (\figurename~\ref{fig:network_architecture}(g)). With dense connectivity, each node in a decoder is presented with not only the final aggregated feature maps but also with the intermediate aggregated feature maps and the original same-scale feature maps from the encoder. As such, the aggregation layer in the decoder node may learn to use only the same-scale encoder feature maps or use all collected feature maps available at the gate.
Unlike U-Net$^e$, deep supervision is not required for UNet+ and UNet++, however, as we will describe later, deep supervision enables model pruning during the inference time, leading to a significant speedup with only modest drop in performance.

\subsection{Technical details}
\label{sec:model}

\vspace{4pt}
\subsubsection{Network connectivity}
Let $x^{i,j}$ denote the output of node X$^{i,j}$ where $i$ indexes the down-sampling layer along the encoder and $j$ indexes the convolution layer of the dense block along the skip connection. The stack of feature maps represented by $x^{i,j}$ is computed as
\begin{equation}
\footnotesize
\label{eq_unet}
    x^{i,j}=\begin{cases}
      \mathcal{H}\left(\mathcal{D}(x^{i-1,j})\right),  & j=0  \\
      \mathcal{H}\left(\left[\left[x^{i,k}\right]_{k=0}^{j-1}, \mathcal{U}(x^{i+1,j-1}) \right]\right), & j>0  \\
    \end{cases}
\end{equation}
where function $\mathcal{H}(\cdot)$ is a convolution operation followed by an activation function, $\mathcal{D}(\cdot)$ and $\mathcal{U}(\cdot)$ denote a down-sampling layer and an up-sampling layer respectively, and $[$ $]$ denotes the concatenation layer. Basically, as shown in~\figurename~\ref{fig:network_architecture}(g), nodes at level $j=0$  receive only one input from the previous layer of the encoder; nodes at level $j=1$ receive two inputs, both from the encoder sub-network but at two consecutive levels; and nodes at level $j>1$ receive $j+1$ inputs, of which $j$ inputs are the outputs of the previous $j$ nodes in the same skip connection and the $j+1^{th}$ input is the up-sampled output from the lower skip connection. The reason that all prior feature maps accumulate and arrive at the current node is because we make use of a dense convolution block along each skip connection.

\vspace{4pt}
\subsubsection{Deep supervision}
We introduce deep supervision in UNet++. For this purpose, we append a 1$\times$1 convolution with $\mathcal{C}$ kernels followed by a \textit{Sigmoid} activation function to the outputs from nodes  X$^{0,1}$, X$^{0,2}$, X$^{0,3}$, and X$^{0,4}$ where $\mathcal{C}$ is the number of classes observed in the given dataset. We then define a hybrid segmentation loss consisting of pixel-wise cross-entropy loss and soft dice-coefficient loss for each semantic scale. The hybrid loss may take advantages of what both loss functions have to offer: smooth gradient and handling of class imbalance~\cite{milletari2016v,sudre2017generalised}. Mathematically, the hybrid loss is defined as: 
\begin{equation}
\footnotesize
\label{eq:object_function}
\mathcal{L}(Y,P) = -\frac{1}{N}\sum_{c=1}^{\mathcal{C}}\sum_{n=1}^{N}{\left(y_{n,c}\log{p_{n,c}}+\frac{2y_{n,c}p_{n,c}}{y^2_{n,c}+p^2_{n,c}}\right)}
\end{equation}
\noindent where $y_{n,c}\in Y$ and $p_{n,c}\in P$ denote the target labels and predicted probabilities for class $c$ and $n^{th}$ pixel in the batch, $N$ indicates the number of pixels within one batch. The overall loss function for UNet++ is then defined as the weighted summation of the hybrid loss from each individual decoders: $\mathcal{L}=\sum_{i=1}^{d}{\eta_i\cdot \mathcal{L}(Y,P^i)}$, where $d$ indexes the decoder. In the experiments, we give same balanced weights $\eta_i$ to each loss, \ie $\eta_i\equiv 1$, and do not process the ground truth for different outputs supervision like Gaussian blur.

\vspace{4pt}
\subsubsection{Model pruning}
Deep supervision enables model pruning. Owing to deep supervision, UNet++ can be deployed in two operation modes: 1) ensemble mode where the segmentation results from all segmentation branches are collected and then averaged, and 2) pruned mode where the segmentation output is selected from only one of the segmentation branches, the choice of which determines the extent of model pruning and speed gain.  \figurename~\ref{fig:prune_structure} shows how the choice of the segmentation branch results in pruned architectures of varying complexity. Specifically, taking the segmentation result from X$^{0,4}$ leads to no pruning whereas taking the segmentation result from X$^{0,1}$ leads to maximal pruning of the network.

\section{Experiments}
\label{sec:experiments}

%##############################################################################################
\begin{table}[t]
\footnotesize
\centering
\caption{Summary of biomedical image segmentation datasets used in our experiments (see \sectionname~\ref{sec:dataset} for details).}
\label{tab:dataset} %
\begin{tabular}{P{0.19\linewidth}P{0.09\linewidth}P{0.15\linewidth}P{0.15\linewidth}P{0.18\linewidth}}
    \hline
    \textbf{Application} & \textbf{Images} & \textbf{Input Size} & \textbf{Modality} & \textbf{Provider} \\
    \hline
    EM & 30 & 96$\times$96 &  microscopy & \tiny{\href{https://imagej.net/Segmentation_of_neuronal_structures_in_EM_stacks_challenge_-_ISBI_2012}{ISBI 2012}}~\cite{cardona2010integrated} \\
    Cell & 354 & 96$\times$96 & Cell-CT & \tiny{\href{https://onlinelibrary.wiley.com/doi/full/10.1002/cncy.21576}{VisionGate}}~\cite{meyer2015cell} \\
    Nuclei & 670 & 96$\times$96 & mixed & \tiny{\href{https://www.kaggle.com/c/data-science-bowl-2018}{Data Science Bowl}} \\
    Brain Tumor & 66,348 & 256$\times$256 & MRI & \tiny{\href{https://www.smir.ch/BRATS/Start2013}{BraTS 2013}}~\cite{menze2015multimodal} \\
    Liver & 331  & 96$\times$96 & CT & \tiny{\href{https://competitions.codalab.org/competitions/17094}{MICCAI 2017 LiTS}} \\
    Lung Nodule & 1,012 & 64$\times$64$\times$64 & CT & \tiny{\href{https://wiki.cancerimagingarchive.net/display/Public/LIDC-IDRI}{LIDC-IDRI}}~\cite{armato2011lung}\\
    \hline
\end{tabular}
\end{table}
%##############################################################################################

%##############################################################################################
\begin{figure}[t]
\begin{center}
\includegraphics[width=1.0\linewidth]{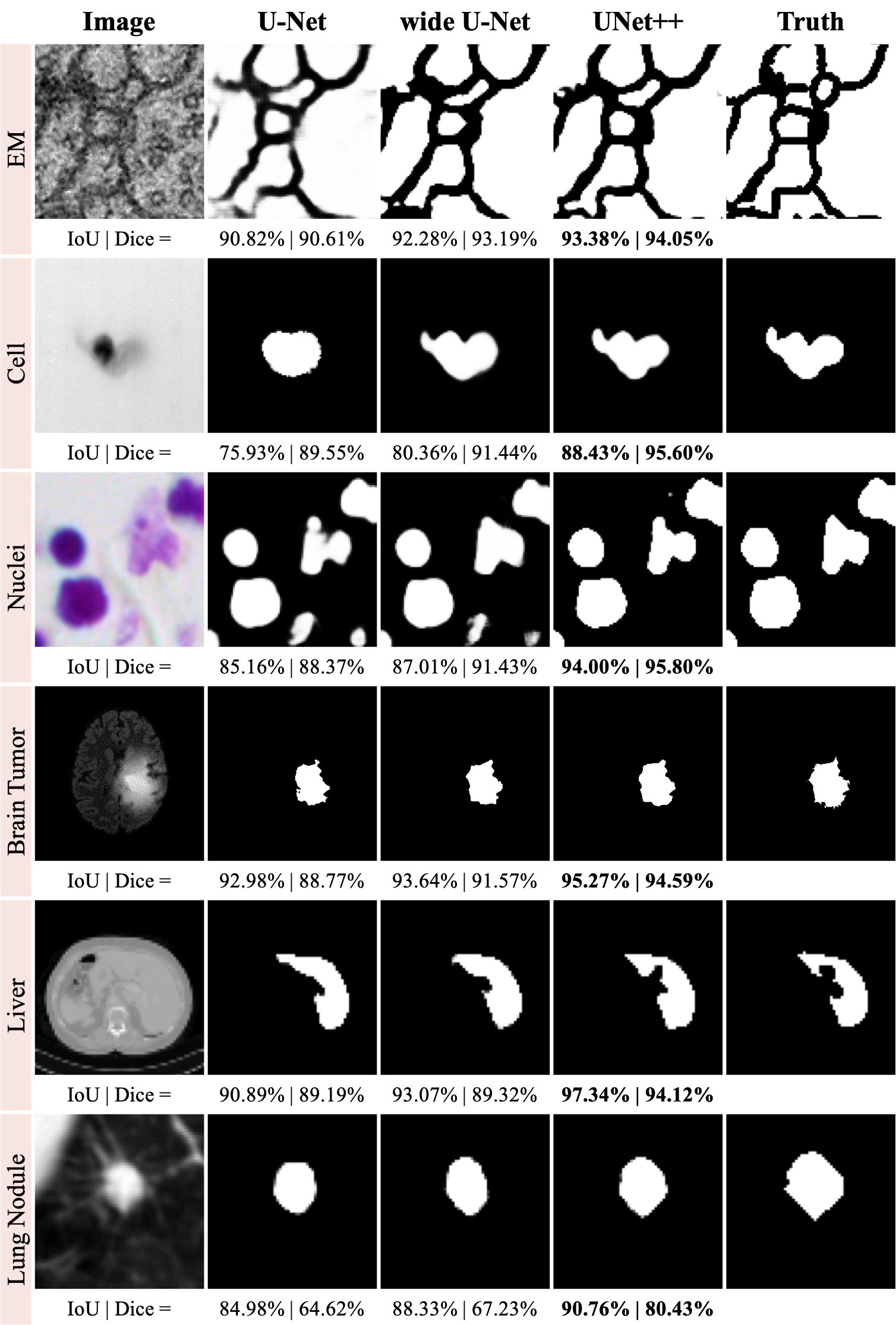}
\end{center}
\caption{Qualitative comparison among U-Net, wide U-Net, and UNet++; showing segmentation results for our six distinct biomedical image segmentation applications. They include various 2D and 3D modalities. The corresponding quantitative scores are provided at the bottom of each prediction (IoU $|$ Dice).}
\label{fig:predict_visualization}
\end{figure}
%##############################################################################################

\subsection{Datasets}
\label{sec:dataset}

\tableautorefname~\ref{tab:dataset} summarizes the six biomedical image segmentation datasets used in this study, covering lesions/organs from most commonly used medical imaging modalities including microscopy, computed tomography (CT), and magnetic resonance imaging (MRI).

\vspace{4pt}
\subsubsection{\texttt{Electron Microscopy (EM)}} The dataset is provided by the EM segmentation challenge~\cite{cardona2010integrated} as a part of ISBI 2012. The dataset consists of 30 images (512$\times$512 pixels) from serial section transmission electron microscopy of the Drosophila firt instar larva ventral nerve cord (VNC).
Referring to the example in~\figurename~\ref{fig:predict_visualization}, each image comes with a corresponding fully annotated ground truth segmentation map for cells (white) and membranes (black). The labeled images are split into training (24 images), validation (3 images), and test (3 images) datasets. 
Both training and inference are done based on 96$\times$96 patches, which are chosen to overlap by half of the patch size via sliding windows. Specifically, during the inference, we aggregate predictions across patches by voting in the overlapping areas.

\vspace{4pt}
\subsubsection{\texttt{Cell}} The dataset is acquired with a Cell-CT imaging system~\cite{meyer2015cell}. Two trained experts manually segment the collected images, so each image in the dataset comes with two binary cell masks. For our experiments, we select a subset of 354 images that have the highest level of agreement between the two expert annotators. The selected images are then split into training (212 images), validation (70 images), and test (72 images) subsets.

\vspace{4pt}
\subsubsection{\texttt{Nuclei}} The dataset is provided by the Data Science Bowl 2018 segmentation challenge and consists of 670 segmented nuclei images from different modalities (brightfield vs. fluorescence). This is the only dataset used in this work with instance-level annotation where each nucleolus is marked in a different color. Images are randomly assigned into a training set (50\%), a validation set (20\%), and a test set (30\%). We then use a sliding window mechanism to extract 96$\times$96 patches from the images, with 32-pixel stride for training and validating model, and with 1-pixel stride for testing. 

\vspace{4pt}
\subsubsection{\texttt{Brain Tumor}} The dataset is provided by BraTS 2013~\cite{menze2015multimodal,kistler2013virtual}. 
To ease the comparison with other approaches, the models are trained using 20 High-grade (HG) and 10  Low-grade (LG) with Flair, T1, T1c, and T2 scans of MR images from all patients, resulting in a total of 66,348 slices.
We further pre-process the dataset by re-scaling the slices to 256$\times$256.
Finally, the  30 patients available in the dataset are randomly assigned into five folds, each having images from six patients. 
We then randomly assign these five folds into a training set (3-fold), a validation set (1-fold), and a test set (1-fold). 
The  ground truth segmentation have four different labels: necrosis, edema, non-enhancing tumor, and enhancing tumor. Following the BraTS 2013, the ``complete'' evaluation is done by considering all four labels as positive class and others as negative class.

\vspace{4pt}
\subsubsection{\texttt{Liver}} The dataset is provided by MICCAI 2017 LiTS Challenge and consists of 331 CT scans, which we split into training (100 patients), validation (15 patients), and test (15 patients) subsets. The ground truth segmentation provides two different labels: liver and lesion. For our experiments, we only consider liver as positive class and others as negative class.

\vspace{4pt}
\subsubsection{\texttt{Lung Nodule}} The dataset is provided by the Lung Image Database Consortium image collection (LIDC-IDRI)~\cite{armato2011lung} and consists of 1018 cases collected by seven academic centers and eight medical imaging companies. Six cases with ground truth issues were identified and removed. The remaining cases were split into training (510), validation (100), and test (408) sets. Each case is a 3D CT scan and the nodules have been marked as volumetric binary masks. We have re-sampled the volumes to 1-1-1 spacing and then extracted a 64$\times$64$\times$64 crop around each nodule. These 3D crops are used for model training and evaluation.

%##############################################################################################
\begin{table}[t]
\footnotesize
\centering
\caption{Details of the architectures used in our study. Wider version of U-Net and V-Net are designed to have comparable number of parameters to UNet++ and VNet++.\iffalse The numbers of convolutional kernels used in 2D and 3D architectures. To show that the performance improvements of our UNet++ (2D) over baseline U-Net and VNet++ (3D) over baseline V-Net architecture are not related to increased parameters rather our architecture's design, we have developed wider version of U-Net and V-Net for comparison having parameters slightly over our UNet++ and VNet++ respectively. We only increase the numbers of convolution kernels in wide architectures while keeping the rest structure exactly same as their baselines.\fi}
\label{tab:wide-unet} %
\begin{tabular}{P{0.17\linewidth}P{0.09\linewidth}P{0.08\linewidth}P{0.08\linewidth}P{0.08\linewidth}P{0.08\linewidth}P{0.08\linewidth}}
    \hline
    \multirow{2}*{\textbf{Architecture}} & \multirow{2}*{\textbf{Params}} & \textbf{X\boldsymbol{$^{0,0}$}} & \textbf{X\boldsymbol{$^{1,0}$}} & \textbf{X\boldsymbol{$^{2,0}$}} & \textbf{X\boldsymbol{$^{3,0}$}} & \textbf{X\boldsymbol{$^{4,0}$}} \\
     &  & \textbf{X\boldsymbol{$^{0,4}$}} & \textbf{X\boldsymbol{$^{1,3}$}} & \textbf{X\boldsymbol{$^{2,2}$}} & \textbf{X\boldsymbol{$^{3,1}$}} & \textbf{X\boldsymbol{$^{4,0}$}} \\
    \hline
    U-Net & 7.8M &32 & 64 & 128 & 256 & 512 \\
    wide U-Net & 9.1M & 35 & 70 & 140 & 280 & 560 \\
    V-Net & 22.6M & 32 & 64 & 128 & 256 & 512 \\
    wide V-Net & 27.0M & 35 & 70 & 140 & 280 & 560 \\
    \hline
    \textbf{Architecture} & \textbf{Params} & \textbf{X\boldsymbol{$^{0,0-4}$}} & \textbf{X\boldsymbol{$^{1,0-3}$}} & \textbf{X\boldsymbol{$^{2,0-2}$}} & \textbf{X\boldsymbol{$^{3,0-1}$}} & \textbf{X\boldsymbol{$^{4,0}$}} \\
    \hline
    UNet+ & 8.7M &32 & 64 & 128 & 256 & 512 \\
    UNet++ & 9.0M &32 & 64 & 128 & 256 & 512 \\
    VNet+ & 25.3M & 32 & 64 & 128 & 256 & 512 \\
    VNet++ & 26.2M & 32 & 64 & 128 & 256 & 512 \\
    \hline
\end{tabular}
\end{table}
%##############################################################################################

\subsection{Baselines and implementation}
\label{sec:implementation}

For comparison, we use the original U-Net~\cite{ronneberger2015u} and a customized wide U-Net architecture for 2D segmentation tasks, and V-Net~\cite{milletari2016v} and a customized wide V-Net architecture for 3D segmentation tasks. We choose U-Net (or V-Net for 3D) because it is a common performance baseline for image segmentation. We have also designed a wide U-Net (or wide V-Net for 3D) with similar number of parameters to our suggested architecture. This is to ensure that the performance gain yielded by our architecture is \textit{not} simply due to increased number of parameters. \tableautorefname~\ref{tab:wide-unet} details the U-Net and wide U-Net architectures. We have further compared the performance of UNet++ against UNet+, which is our intermediate architecture proposal. The numbers of kernels in the intermediate nodes have been given in~\tableautorefname~\ref{tab:wide-unet}. 

Our experiments are implemented in Keras with Tensorflow backend. We use {\em early-stop} mechanism on the validation set to avoid over-fitting and evaluate the results using Dice-coefficient and Intersection over Union (IoU). Alternative measurement metrics, such as pixel-wise sensitivity, specificity, F1, and F2 scores, along with the statistical analysis can be found in Appendix~\sectionname~\ref{sec:various_meansures}. Adam is used as the optimizer with a learning rate of 3$e$-4. Both UNet+ and UNet++ are constructed from the original U-Net architecture. All the experiments are performed using three NVIDIA TITAN X (Pascal) GPUs with 12 GB memory each.

%##############################################################################################

\begin{table*}[t]
\footnotesize
\begin{center}
\begin{threeparttable}
\caption{{\jlrev Semantic segmentation results measured by IoU (mean$\pm$s.d. \%) for U-Net, wide U-Net, UNet+ (our intermediate proposal), and  UNet++ (our final proposal). Both UNet+ and UNet++ are evaluated with and without deep supervision (DS). We have performed independent two sample $t$-test between U-Net~\cite{falk2018u} vs. others for 20 independent trials and highlighted boxes in red when the differences are statistically significant ($p<0.05$).}}
\label{tab:main_results}
{\jlrev
\begin{tabular}{P{0.074\linewidth}P{0.01\linewidth}P{0.05\linewidth}|P{0.06\linewidth}P{0.06\linewidth}P{0.06\linewidth}P{0.1\linewidth}P{0.06\linewidth}|P{0.074\linewidth}P{0.01\linewidth}P{0.05\linewidth}|P{0.11\linewidth}}
    \hline
    \multirow{2}*{\textbf{Architecture}} & \multirow{2}*{\textbf{DS}} & \multirow{2}*{\textbf{Params}} & \multicolumn{5}{c|}{\textbf{2D Application}} & \multirow{2}*{\textbf{Architecture}} & \multirow{2}*{\textbf{DS}} & \multirow{2}*{\textbf{Params}} & \textbf{3D Application} \\
    \cline{4-8}\cline{12-12}
     & & &  \textbf{EM} & \textbf{Cell} & \textbf{Nuclei} & \textbf{Brain Tumor$^{\dagger}$} & \textbf{Liver} & & & & \textbf{Lung Nodule} \\
    \hline
    U-Net~\cite{falk2018u} & \xmark & 7.8M & 88.30{\tiny $\pm$0.24} & 88.73{\tiny $\pm$1.64} & 90.57{\tiny $\pm$1.26} & 89.21{\tiny $\pm$1.55} & 79.90{\tiny $\pm$1.38} & V-Net~\cite{milletari2016v} & \xmark & 22.6M & 71.17{\tiny $\pm$4.53} \\
    wide U-Net & \xmark & 9.1M & 88.37{\tiny $\pm$0.13} & 88.91{\tiny $\pm$1.43} & 90.47{\tiny $\pm$1.15} & 89.35{\tiny $\pm$1.49}  & 80.25{\tiny $\pm$1.31} & wide V-Net & \xmark & 27.0M & 73.12{\tiny $\pm$3.99} \\
    UNet+ & \xmark & 8.7M & 88.39{\tiny $\pm$0.15} & \cellcolor{maroon!15}90.71{\tiny $\pm$1.25} & \cellcolor{maroon!15}91.73{\tiny $\pm$1.09} & \cellcolor{maroon!15}90.70{\tiny $\pm$0.91} & 79.62{\tiny $\pm$1.20} & VNet+ & \xmark & 25.3M & \cellcolor{maroon!15}75.93{\tiny $\pm$2.93} \\
    UNet+ & \cmark & 8.7M & \cellcolor{maroon!15}88.89{\tiny $\pm$0.12} & \cellcolor{maroon!15}91.18{\tiny $\pm$1.13} & \cellcolor{maroon!15}92.04{\tiny $\pm$0.89} & \cellcolor{maroon!15}91.15{\tiny $\pm$0.65} & \cellcolor{maroon!15}\textbf{82.83{\tiny $\pm$0.92}} & VNet+ & \cmark & 25.3M & \cellcolor{maroon!15}76.72{\tiny $\pm$2.48} \\
    UNet++ & \xmark & 9.0M & \cellcolor{maroon!15}88.92{\tiny $\pm$0.14} & \cellcolor{maroon!15}91.03{\tiny $\pm$1.34} &  \cellcolor{maroon!15}\textbf{92.44{\tiny $\pm$1.20}} & \cellcolor{maroon!15}90.86{\tiny $\pm$0.81}&	\cellcolor{maroon!15}82.51{\tiny $\pm$1.29} & VNet++ & \xmark & 26.2M & \cellcolor{maroon!15}76.24{\tiny $\pm$3.11}\\
    UNet++ & \cmark & 9.0M& \cellcolor{maroon!15}\textbf{89.33{\tiny $\pm$0.10}} & \cellcolor{maroon!15}\textbf{91.21{\tiny $\pm$0.98}}  & \cellcolor{maroon!15}92.37{\tiny $\pm$0.98} & \cellcolor{maroon!15}\textbf{91.21{\tiny $\pm$0.68}}	& \cellcolor{maroon!15}82.60{\tiny $\pm$1.11} & VNet++ & \cmark & 26.2M & \cellcolor{maroon!15}\textbf{77.05{\tiny $\pm$2.42}}\\
    \hline
    \end{tabular}
    \begin{tablenotes}
        \scriptsize
        \item $^{\dagger}$ The winner in {\href{https://www.smir.ch/BRATS/Start2013#trainingResults}{BraTS 2013} holds a ``complete'' Dice of 92$\%$  vs. 90.83$\%\pm$2.46$\%$ (our UNet++ with deep supervision).}
    \end{tablenotes}
}
\end{threeparttable}
\end{center}
\end{table*}
%##############################################################################################

%##############################################################################################

\begin{figure*}[t]
\begin{center}
\includegraphics[width=1.0\linewidth]{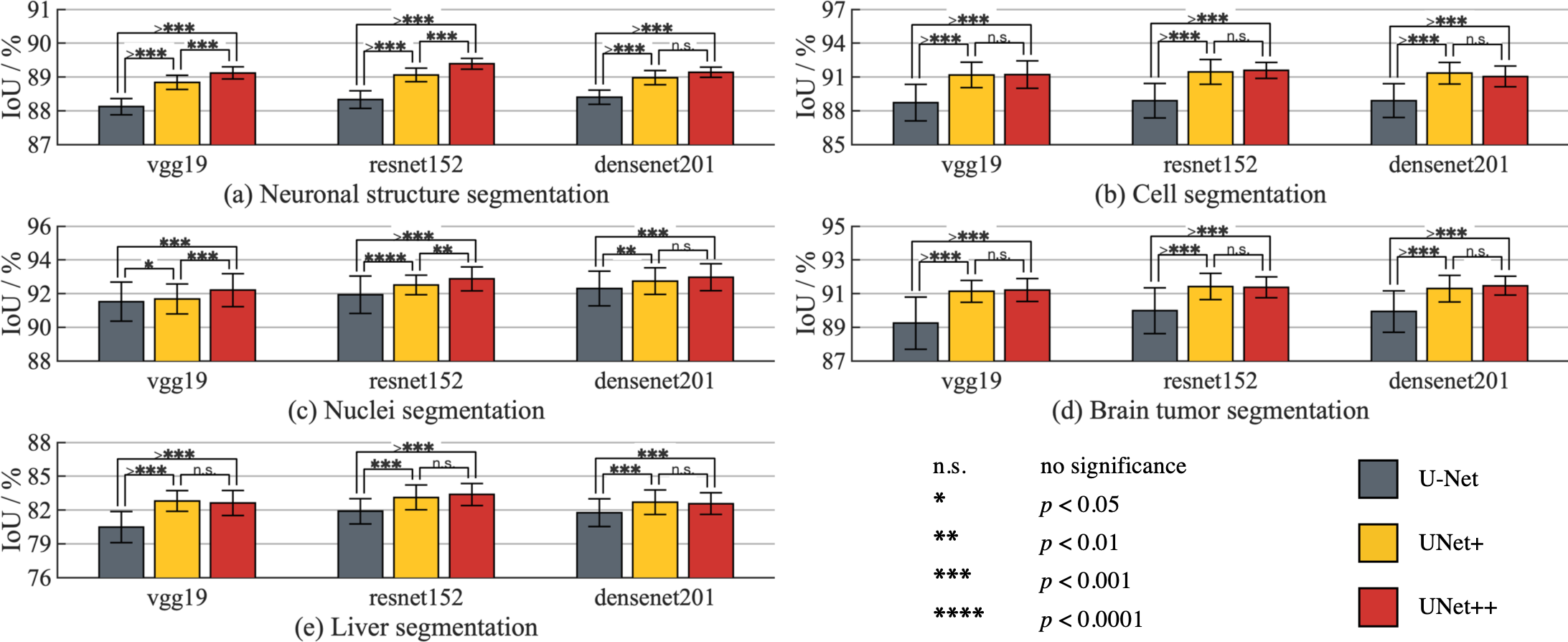}
\end{center}
\caption{\jlrev{Comparison between U-Net, UNet+, and UNet++ when applied to the state-of-the-art backbones for the tasks of neuronal structure, cell, nuclei, brain tumor, and liver segmentation. UNet++, trained with deep supervision, consistently outperforms U-Net across all backbone architectures and applications under study. By densely connecting the intermediate layers, UNet++ also yields higher segmentation performance than UNet+ in most experimental configurations.
The error bars represent the 95\% confidence interval and the number of $\ast$ on the bridge indicates the level of significance measured by $p$-value (``n.s.'' stands for ``not statistically significant'').}}
\label{fig:various_backbones}
\end{figure*}
%##############################################################################################

\section{Results}
\label{sec:results}
                                                                                                                                                                                 
\subsection{Semantic segmentation results}
\label{sec:semantic_results}

\tableautorefname~\ref{tab:main_results} compares U-Net, wide U-Net, UNet+, and UNet++ in terms of the number parameters and {\jlrev segmentation results measured by IoU (mean$\pm$s.d)} for the six segmentation tasks under study. As seen, wide U-Net consistently outperforms U-Net. This improvement is attributed to the larger number of parameters in wide U-Net. {\jlrev UNet++ without deep supervision achieves a significant IoU gain over both U-Net and wide U-Net for all the six tasks of neuronal structure ($\uparrow$0.62$\pm$0.10, $\uparrow$0.55$\pm$0.01), cell ($\uparrow$2.30$\pm$0.30, $\uparrow$2.12$\pm$0.09), nuclei ($\uparrow$1.87$\pm$0.06, $\uparrow$1.71$\pm$0.06), brain tumor ($\uparrow$2.00$\pm$0.87, $\uparrow$1.86$\pm$0.81), liver ($\uparrow$2.62$\pm$0.09, $\uparrow$2.26$\pm$0.02), and lung nodule ($\uparrow$5.06$\pm$1.42, $\uparrow$3.12$\pm$0.88) segmentation.} Using deep supervision and average voting further improves UNet++, increasing the IoU by up to 0.8 points. Specifically, neuronal structure and lung nodule segmentation benefit the most from deep supervision because they appear at varying scales in EM and CT slices.
Deep supervision, however, is only marginally effective for other datasets at best.  \figurename~\ref{fig:predict_visualization} depicts a qualitative comparison between the results of U-Net, wide U-Net, and UNet++. 

{\jlrev We have further investigated the extensibility of UNet++ for semantic segmentation by applying redesigned skip connections to an array of modern CNN architectures: vgg-19~\cite{simonyan2014very}, resnet-152~\cite{he2016deep}, and densenet-201~\cite{huang2017densely}.} Specifically, we have turned each architecture above into a U-Net model by adding a decoder sub-network, and then replaced the plain skip connections of U-Net with the redesigned connections of UNet++. For comparison, we have also trained U-Net and  UNet+ with the aforementioned backbone architectures. 
For a comprehensive comparison, we have used \texttt{EM}, \texttt{Cell}, \texttt{Nuclei}, \texttt{Brain Tumor} and \texttt{Liver} segmentation datasets.
As seen in \figurename~\ref{fig:various_backbones}, UNet++ consistently outperforms U-Net and UNet+ across all backbone architectures and applications under study.
Through 20 trials, we further present statistical analysis based on the independent two-sample $t$-test on each pair among U-Net, UNet+, and UNet++.
Our results suggest that UNet++ is an effective, backbone-agnostic extension to U-Net. To facilitate reproducibility and model reuse, we have released the implementation\footnote{The project page: \href{https://github.com/MrGiovanni/UNetPlusPlus}{https://github.com/MrGiovanni/UNetPlusPlus}
} of U-Net, UNet+, and UNet++ for various traditional and modern backbone architectures. 

%##############################################################################################
\begin{table}[t]
\footnotesize
\begin{center}
\begin{threeparttable}
\caption{Redesigned skip connections improve both semantic and instance segmentation for the task of nuclei segmentation. We use Mask R-CNN for instance segmentation and U-Net for semantic segmentation in this comparison.}
\label{tab:fcn_family_performance}

\begin{tabular}{P{0.27\linewidth}P{0.13\linewidth}P{0.12\linewidth}P{0.12\linewidth}P{0.12\linewidth}}
    \hline
     \textbf{Architecture} & \textbf{Backbone} & \textbf{IoU} & \textbf{Dice} & \textbf{Score} \\
    \hline
    U-Net &resnet101&91.03 & 75.73 & 0.244 \\
    \rowcolor{maroon!15}
     UNet++&resnet101&   \bd{92.55} & \bd{89.74} & \bd{0.327} \\
    \hline
    Mask R-CNN~\cite{he2017mask} &resnet101& 93.28 & 87.91 & 0.401 \\
    \rowcolor{maroon!15}
    Mask RCNN++$^{\dagger}$&resnet101&\bd{95.10} & \bd{91.36} & \bd{0.414} \\
    \hline
    \multicolumn{5}{l}{$^{\dagger}$Mask R-CNN with UNet++ design in its feature pyramid.}
        \end{tabular}
\end{threeparttable}
\end{center}
\end{table}
%##############################################################################################

\subsection{Instance segmentation results}
\label{sec:instance_segmentation}

Instance segmentation consists in segmenting and distinguishing all object instances; hence, more challenging than semantic segmentation. We use Mask R-CNN~\cite{he2017mask} as the baseline model for instance segmentation. Mask R-CNN utilizes feature pyramid network (FPN) as backbone to generate object proposal at multiple scales, and then outputs the segmentation masks for the collected proposals via a dedicated segmentation branch. We modify Mask R-CNN by replacing the plain skip connections of FPN with the redesigned skip connections of UNet++. We refer to this model as Mask RCNN++. We use resnet101 as the backbone for Mask R-CNN in our experiments. 

\tableautorefname~\ref{tab:fcn_family_performance} compares the performance of Mask R-CNN and Mask RCNN++ for nuclei segmentation. 
{\jlrev We have chosen the \texttt{Nuclei} dataset because multiple nucleolus instances can be present in an image, in which case each instance is annotated in a different color, and thus marked as a distinct object. Therefore, this dataset is amenable to both semantic segmentation where all nuclei instances are treated as foreground class, and also instance segmentation where each individual nucleus is to be segmented separately.}
As seen in \tableautorefname~\ref{tab:fcn_family_performance}, Mask RCNN++ outperforms its original counterpart, achieving 1.82 points increase in IoU (93.28\% to 95.10\%), 3.45 points increase in Dice (87.91\% to 91.36\%), and 0.013 points increase in the leaderboard score (0.401 to 0.414). To put this performance in perspective, we have also trained a U-Net and UNet++ model for semantic segmentation with a resnet101 backbone. As seen in \tableautorefname~\ref{tab:fcn_family_performance}, Mask R-CNN models achieve higher segmentation performance than semantic segmentation models. Furthermore, as expected, UNet++ outperforms U-Net for semantic segmentation.

%##############################################################################################
\begin{figure}[t]
\begin{center}
\includegraphics[width=1.0\linewidth]{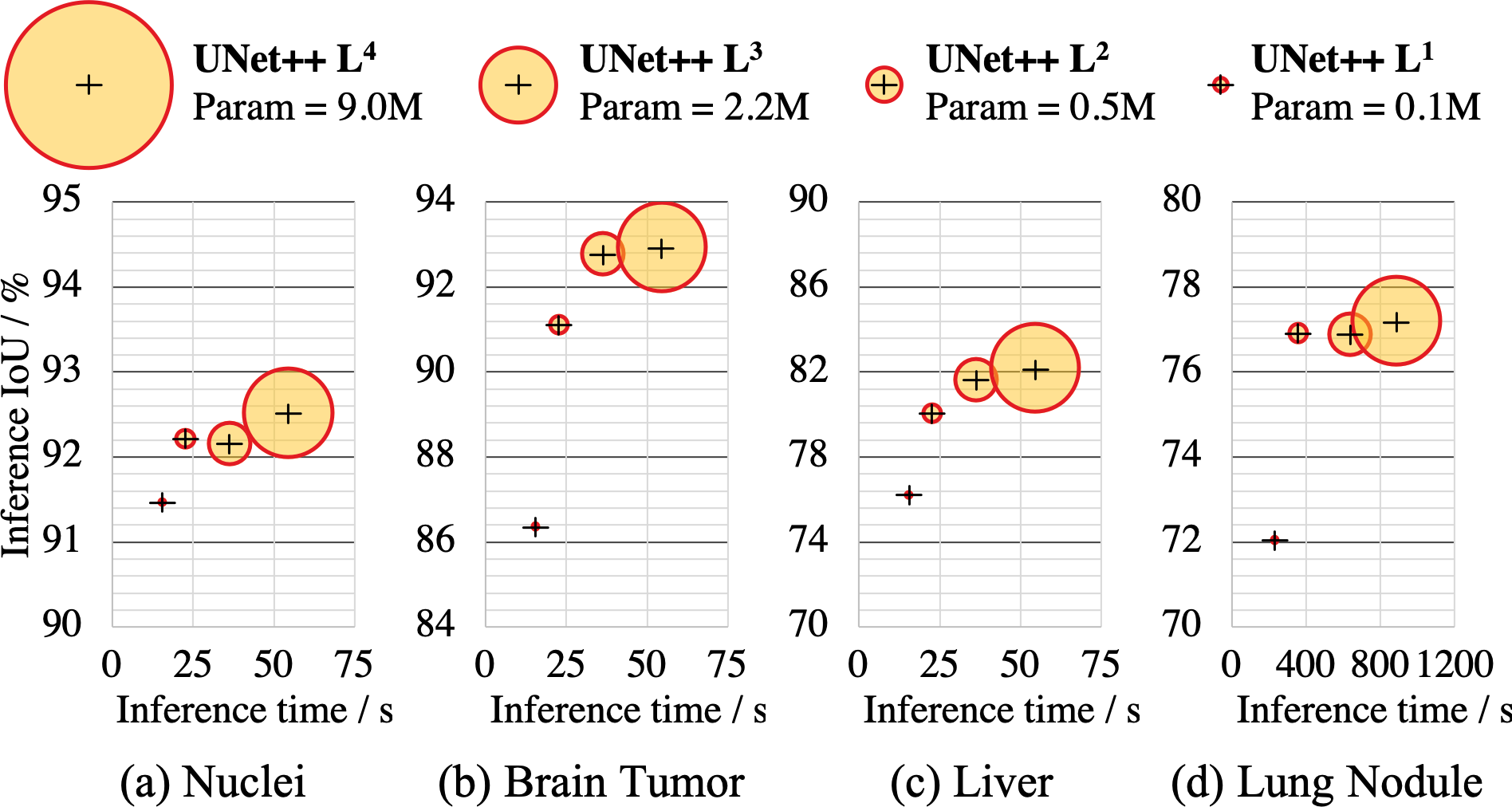}
\end{center}
\caption{Complexity (size $\propto$ parameters), inference time, and IoU of UNet++ under different levels of pruning. \iffalse (see \sectionname~\ref{sec:model_pruning}).\fi The inference time is calculated by the time taken to process 10K test images on a single NVIDIA TITAN X (Pascal) GPU with 12 GB memory.}
\label{fig:speed_accuracy}
\end{figure}
%##############################################################################################

\subsection{Model pruning}
\label{sec:model_pruning}

Once UNet++ is trained, the decoder path for depth $d$ at inference time is completely independent from the decoder path for depth $d+1$. As a result, we can completely remove the decoder for depth $d+1$, obtaining a shallower version of the trained UNet++ at depth $d$, owing to the introduced deep supervision. This pruning can significantly reduce the inference time, but segmentation performance may degrade. As such, the level of pruning should be determined by evaluating the model's performance on the validation set. We have studied the inference speed-IoU trade-off for UNet++ in \figurename~\ref{fig:speed_accuracy}. We use UNet++ L$^{d}$ to denote UNet++ pruned at depth $d$ (see \figurename~\ref{fig:prune_structure} for further details). As seen, UNet++ L$^{3}$ achieves on average 32.2\% reduction in inference time and 75.6\% reduction in memory footprint while degrading IoU by only 0.6 points. More aggressive pruning further reduces the inference time but at the cost of significant IoU degradation. More importantly, this observation has the potential to exert important impact on computer-aided diagnosis (CAD) on mobile devices, as the existing deep convolutional neural network models are computationally expensive and memory intensive.

%##############################################################################################
\begin{figure}[t]
\begin{center}
\includegraphics[width=1.0\linewidth]{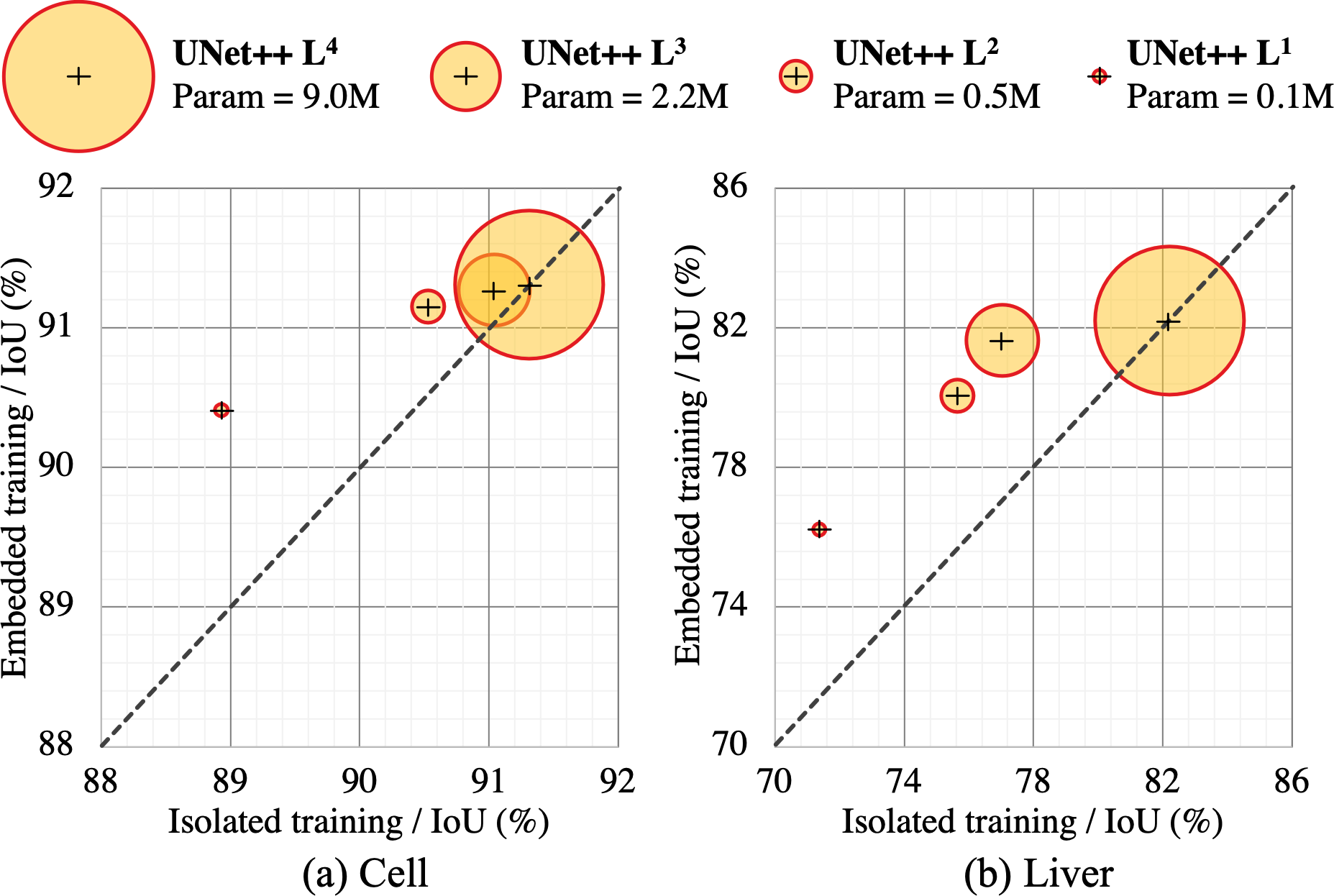}
\end{center}
\caption{We demonstrate that our architectural design improves the performance of each shallower network embedded in UNet++. The embedded shallower networks show improved segmentation when pruned from UNet++ in comparison to the same network trained isolated. Due to no pruning, UNet++ L$^4$ naturally achieves the same level of performance in isolated and embedded training modes.}
\label{fig:pruned_vs_stand-alone}
\end{figure}
%##############################################################################################

\subsection{Embedded vs. isolated training of pruned models}
\label{sec:embedded_vs_isolated_training}

In theory, UNet++ L$^d$ can be trained in two fashions: 1) embedded training where the full UNet++ model is trained and then pruned at depth $d$ to obtain UNet++ L$^d$, 2) isolated training where UNet++ L$^d$ is trained in isolation without any interactions with the deeper encoder and decoder nodes. Referring to~\figurename~\ref{fig:prune_structure}, embedded training of a sub-network consists of training all graph nodes (both yellow and grey components) {\jlrev with deep supervision}, but we then use only the yellow sub-network during the inference time. In contrast, isolated training consists of removing the grey nodes from the graph, basing the training and test solely on the yellow sub-network.

We have compared the isolated and embedded training schemes for various levels of UNet++ pruning across two datasets in~\figurename~\ref{fig:pruned_vs_stand-alone}. We have discovered that the embedded training of UNet++ L$^d$ results in a higher performing model than training the same architecture in isolation. The observed superiority is more pronounced under aggressive pruning when the full UNet++ is pruned to UNet++ L$^1$. In particular, the embedded training of UNet++ L$^1$ for liver segmentation achieves 5-point increase in IoU over the isolated training scheme. This finding suggests that supervision signal coming from the deep downstream enables training higher performing shallower models. This finding is also related to knowledge distillation where the knowledge learned by a deep teacher network is learned by a shallower student network.

%##############################################################################################
\begin{figure}[t]
\begin{center}
\includegraphics[width=1.0\linewidth]{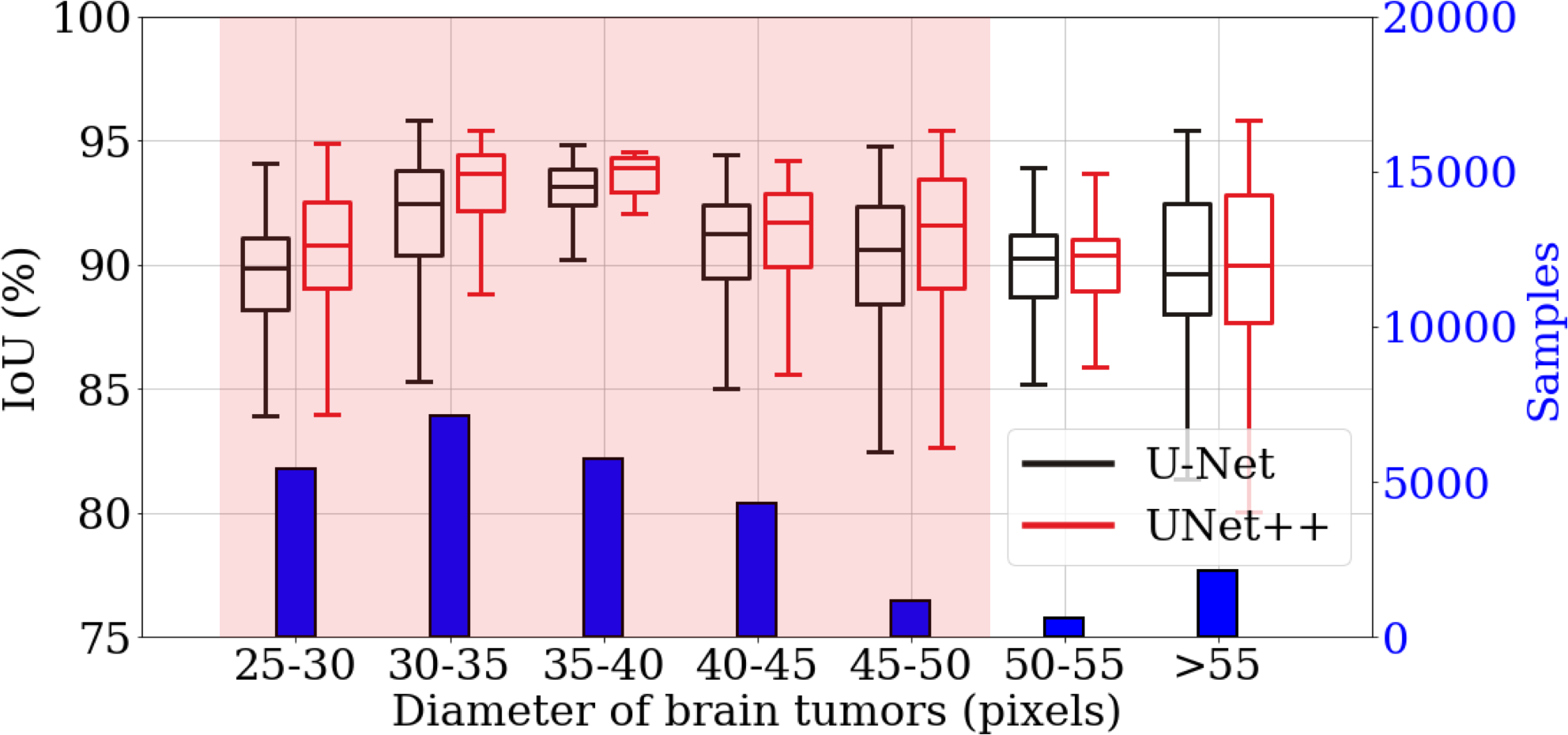}
\end{center}
\caption{UNet++ can better segment tumors of various sizes than does U-Net. We measure the size of tumors based on the ground truth masks and then divide them into seven groups. The histogram shows the distribution of different tumor sizes. The box-plot compares the segmentation performances of U-Net (black) and UNet++ (red) in each group. The $t$-test for two independent samples has been further performed on each group. As seen, UNet++ improves segmentation for all sizes of tumors and the improvement is significant ($p<0.05$) for the majority of the tumor sizes (highlighted in red).}
\label{fig:multi_depth_improvement_brain_tumor}
\end{figure}
%##############################################################################################

%##############################################################################################
\begin{figure*}[t]
\begin{center}
\includegraphics[width=1.0\linewidth]{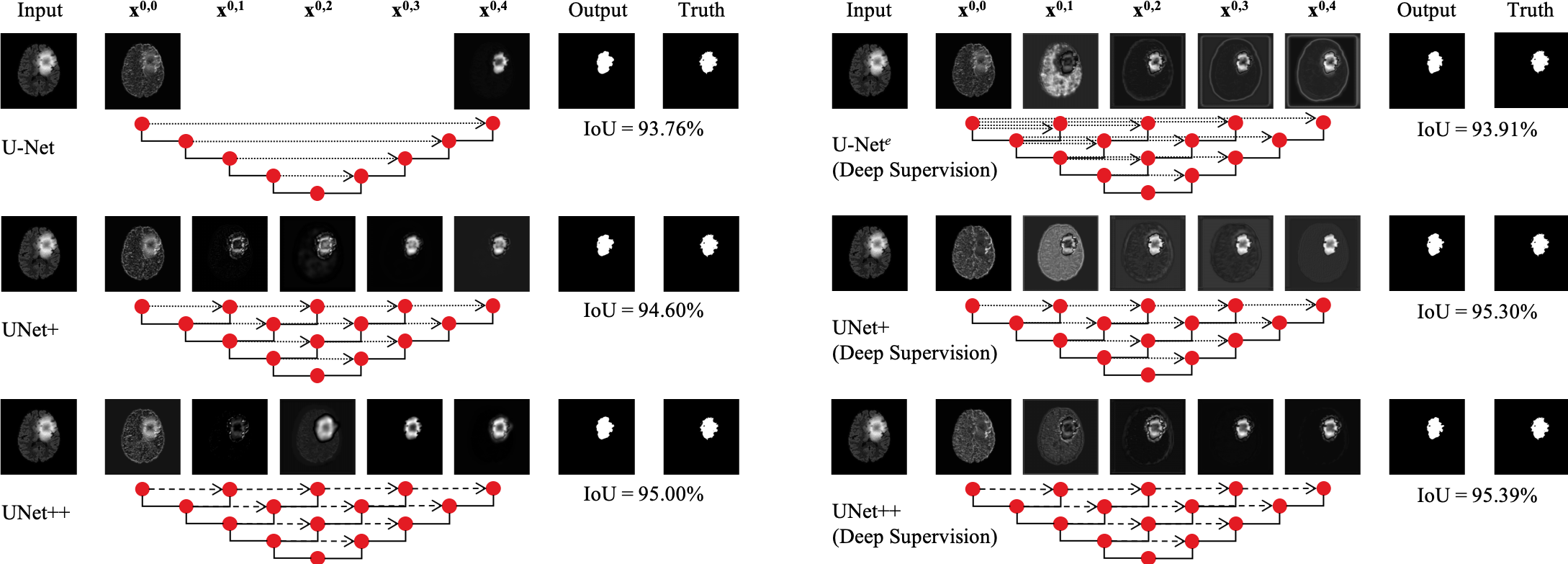}
\end{center}
\caption{Visualization and comparison of feature maps from early, intermediate, and late layers along the top most skip connection for brain tumor images. \iffalse (see \sectionname~\ref{sec:feature_aggregation}).\fi Here, the dot arrows denote plain skip connection in U-Net and UNet+, while the dash arrows denote dense connections introduced in UNet++.}
\label{fig:feature_map_aggregation}
\end{figure*}
%##############################################################################################

\section{Discussions}
\label{sec:discussion}
{\jlrevrev
\subsection{Performance analysis on stratified lesion sizes}
\label{sec:receptive}

\figurename~\ref{fig:multi_depth_improvement_brain_tumor} compares U-Net and UNet++ for segmenting different sizes of brain tumors. To avoid clutter in the figure, we group the tumors by size into seven buckets. As seen, UNet++ consistently outperforms U-Net across all the buckets. We also adopt $t$-test on each bucket based on 20 different trials to measure the significance of the improvement, concluding that 5 out of the 7 comparisons are statistically significant ($p < 0.05$). The capability of UNet++ in segmenting tumors of varying sizes is attributed to its built-in ensemble of U-Nets, which enables image segmentation based on multi-receptive field networks.}

\subsection{Feature maps visualization}
\label{sec:feature_aggregation}

In Section~\ref{sec:motiv}, we explained that the redesigned skip connections enable the fusion of semantically rich decoder feature maps with feature maps of varying semantic scales from the intermediate layers of the architecture. In this section, we illustrate this privilege of our re-designed skip connections by visualizing the intermediate feature maps.

\figurename~\ref{fig:feature_map_aggregation} shows representative feature maps from early, intermediate, and late layers along the top most skip connection (\ie X$^{0,i}$) for a brain tumor image. The representative feature map for a layer is obtained by averaging all its feature maps. Also note that architectures in the left side of \figurename~\ref{fig:feature_map_aggregation} are trained using only loss function appended to the deepest decoder layer (X$^{0,4}$) whereas the architectures in the right side of \figurename~\ref{fig:feature_map_aggregation} are trained with deep supervision. 
{\jlrev Note that these feature maps are not the final outputs. We have appended an additional 1$\times$1 convolutional layer on top of each decoder branch to form the final segmentation.}
We observe that the outputs of U-Net's intermediate layers are semantically dissimilar  whereas for UNet+ and UNet++ the outputs are formed gradually. The output of node X$^{0,0}$ in U-Net undergoes slight transformation (few convolution operations only) whereas the output of X$^{1,3}$, the input of X$^{0,4}$, goes through nearly every transformation (four down-sampling and three up-sampling stages) learned by the network. Hence, there is a large gap between the representation capability of X$^{0,0}$ and X$^{1,3}$. So, simply concatenating the outputs of X$^{0,4}$ and X$^{1,3}$ is not an optimal solution. In contrast, redesigned skip connections in UNet+ and UNet++ help refine the segmentation result gradually. 
{\jlrev We further present the learning curves of all six medical applications in Appendix~\sectionname~\ref{sec:learning_curves}, revealing that the addition of dense connections in UNet++ encourages a better optimization and reaches lower validation loss.}

\subsection{Collaborative learning in UNet++}
\label{sec:collarborative_learning}

Collaborative learning is known as training multiple classifier heads of the same network simultaneously on the same training data. It is found to improve the generalization power of deep neural networks~\cite{song2018collaborative}. UNet++ naturally embodies collaborative learning through aggregating multi-depth networks and supervising segmentation heads from each of the constituent networks. Besides, the segmentation heads, for example X$^{0,2}$ in \figurename~\ref{fig:prune_structure}, receive gradients from both strong (loss from ground truth) and soft (losses propagated from adjacent deeper nodes) supervision. As a result, the shallower networks improve their segmentation (\figurename~\ref{fig:pruned_vs_stand-alone}) and provide more informative representation to deeper counterparts. Basically, deeper and shallower networks regularize each other via collaborative learning in UNet++. Training multi-depth embedded networks together results in improved segmentation than training them individually as isolated network which is evident in \sectionname~\ref{sec:embedded_vs_isolated_training}. {\jlrev The embedded design of UNet++ makes it amenable to auxiliary training, multi-task learning, and knowledge distillation~\cite{bengio2009learning,hinton2015distilling,song2018collaborative}.}

\section{Related Works}
\label{sec:related_works}
In the following, we review the works related to redesigned skip connections, feature aggregation, and deep supervision, which are the main components of our new architecture.

\subsection{Skip connections}
\label{sec:related_work_skip_connections}

Skip connections were first introduced in the seminal work of
Long~\etal~\cite{long2015fully} where they proposed a fully convolutional networks (FCN) for semantic segmentation. Shortly after, building on skip connections, Ronneberger~\etal~\cite{ronneberger2015u} proposed U-Net architecture for semantic segmentation in medical images. The FCN and U-Net architectures however differ in how the up-sampled decoder feature maps were fused with the same-scale feature maps from the encoder network. While  FCN~\cite{long2015fully} uses the summation operation for feature fusion,  U-Net~\cite{ronneberger2015u} concatenates the features followed by the application of convolutions and non-linearities. The skip connections have shown to help recover the full spatial resolution, making fully convolutional methods suitable for semantic segmentation~\cite{chaurasia2017linknet,lin2017refinenet,zhao2018icnet,tajbakhsh2019errornet}. 
Skip connections have further been used in modern neural architectures such as  residual networks~\cite{he2016deep,he2016identity} and dense networks \cite{huang2017densely}, facilitating the gradient flow and improving the overall performance of classification networks.

\subsection{Feature aggregation}
\label{sec:related_work_aggregation}
The exploration of aggregating hierarchical feature has recently been the subject of research. Fourure~\etal~\cite{fourure2017gridnet} propose GridNet, which is an encoder-decoder architecture wherein the feature maps are wired in a grid fashion, generalizing several classical segmentation architectures. Despite GridNet contains multiple streams with different resolutions, it lacks up-sampling layers between skip connections; and thus, it does not represent UNet++.
Full-resolution residual networks (FRRN)~\cite{pohlen2017full} employs a two-stream system, where full-resolution information is carried in one stream and context information in the other pooling stream. In~\cite{jiang2019multiple}, two improved versions of FRRN are proposed, \ie incremental MRRN with 28.6M parameters and dense MRRN with 25.5M parameters. 
{\jlrev These 2D architectures however have similar number of parameters to our 3D VNet++ and three times more parameters than 2D UNet++; and thus, simply upgrading these architectures to a 3D manner may not be amenable to the common 3D volumetric medical imaging applications.}
We would like to note that our redesigned dense skip connections are completely different from those used in MRRN, which consists of a common residual stream. Also, it's not flexible to apply the design of MRRN to other backbone encoders and meta framework such as Mask R-CNN~\cite{he2017mask}. DLA\footnote{Deep Layer Aggregation---a simultaneous but independent work published in CVPR-2018~\cite{yu2018deep}.}~\cite{yu2018deep}, topologically equivalent to our intermediate architecture UNet+ (\figurename~\ref{fig:network_architecture}(f)), sequentially connects the same resolution of feature maps, without long skip connections as used in U-Net. Our experimental results demonstrate that by densely connecting the layers, UNet++ achieves higher segmentation performance than UNet+/DLA (see \tableautorefname~\ref{tab:main_results}). 

\subsection{Deep supervision}
\label{sec:related_work_deep_supervision}

He~\etal~\cite{he2016deep} suggested that the depth $d$ of network can act as a regularizer. Lee~\etal~\cite{lee2015deeply} demonstrated that deeply supervised layers can improve the learning ability of the hidden layer, enforcing the intermediate layers to learn discriminative features, enabling  fast convergence and regularization of the network~\cite{dou20173d}. DenseNet~\cite{huang2017densely} performs a similar deep supervision in an implicit fashion. Deep supervision can be used in U-Net like architecture as well.
Dou~\etal~\cite{dou20163d} introduce a deep supervision by combining predictions from varying resolutions of feature maps, suggesting that it can combat potential optimization difficulties and thus reach faster convergence rate and more powerful discrimination capability. Zhu~\etal~\cite{zhu2017deeply} used eight additional deeply supervised layers in their proposed architecture. 
Our nested networks are however more amenable to training under deep supervision: 1) multiple decoders automatically generate full resolution segmentation maps; 2) the networks are embedded various different depths of U-Net so that it grasps multiple-resolution features; 3) densely connected feature maps help smooth the gradient flow and give relatively consistent predicting mask; 4) the high dimension features have effects on every outputs through back-propagation, allowing us to prune the network in the inference phase.

\subsection{Our previous work}
\label{sec:ours}

We first presented UNet++ in our DLMIA 2018 paper~\cite{zhou2018unet++}. 
{\jlrev UNet++ has since been quickly %widely 
adopted by the research community, either as a strong baseline for comparison~\cite{SunZJCXLMWLW19,fang2019selective,fang2019improved,meng2020multiscale}, or as a source of inspiration for developing newer semantic segmentation architectures~\cite{zhang2018mdu,chen2018improved,zhou2018learning,wu2019automatical,song2019u,yang2019eda}; it has also been utilized for multiple applications, such as segmenting objects in biomedical images~\cite{zyuzin2019comparison,cui2019pulmonary}, natural images~\cite{SunXLW19}, and satellite images~\cite{peng2019end,zhang2019web}. Recently, Shenoy~\cite{shenoyfeature} has independently and systematically investigated UNet++ for the task of ``contact prediction model PconsC4'', demonstrating significant improvement over widely-used U-Net. 
}

Nevertheless, to further strengthen UNet++ on our own, the current work presents several extensions to our previous work:
(1) we present a comprehensive study on network depth, motivating the need for the proposed architecture (\sectionname~\ref{sec:motiv});
(2) we compare the embedded training schemes with the isolated ones at various levels of pruned UNet++, and discover that training embedded U-Nets of multi-depths leads to improved performance than individually training them in isolation (\sectionname~\ref{sec:embedded_vs_isolated_training});
(3) we strengthen our experiments by including a new magnetic resonance imaging (MRI) dataset for brain tumor segmentation (\sectionname~\ref{sec:results});
(4) we demonstrate the effectiveness of UNet++ in Mask R-CNN, resulting in a more powerful model namely Mask RCNN++ (\sectionname~\ref{sec:instance_segmentation});
(5) we investigate the extensibility of UNet++ to multiple advanced encoder backbones for semantic segmentation (\sectionname~\ref{sec:semantic_results});
(6) we study the effectiveness of UNet++ in segmenting lesions of varying sizes (\sectionname~\ref{sec:receptive}); and
(7) we visualize the feature propagation along the resigned skip connection to explain the performance (\sectionname~\ref{sec:feature_aggregation}).

\section{Conclusion}
\label{sec:conclusion}

We have presented a novel architecture, named UNet++, for more accurate image segmentation. 
The improved performance by our UNet++ is attributed to its nested structure and re-designed skip connections, which aim to address two key challenges of the U-Net: 1) unknown depth of the optimal architecture and 2) the unnecessarily restrictive design of skip connections. We have evaluated UNet++ using six distinct biomedical imaging applications and demonstrated consistent performance improvement over various state-of-the-art backbones for semantic segmentation and meta framework for instance segmentation.

\section*{Acknowledgments}
This research has been supported partially by ASU and Mayo Clinic through a Seed Grant and an Innovation Grant, and partially by NIH under Award Number R01HL128785. The content is solely the responsibility of the authors and does not necessarily represent the official views of NIH. We thank M. R. Hosseinzadeh Taher and F. Haghighi for their verification of liver segmentation performance and the ablation study of embedded and isolated UNet++. We also thank Michael G. Meyer for allowing us to test our ideas on the Cell-CT dataset. The content of this paper is covered by US patents pending.

\bibliographystyle{IEEEtran}
\bibliography{refs}

\newpage
\newpage

\appendices

{\jlrev
\section{Additional Measurements}
\label{sec:various_meansures}

\begin{table}[h]
\footnotesize
\begin{center}
\begin{threeparttable}
\caption{{\jlrev Pixel-wise sensitivity, specificity, F1, and F2 scores for all six applications under study. Note that the $p$-values are calculated between our UNet++ with deep supervision vs. the original U-Net.
As seen, powered by redesigned skip connections and deep supervision, UNet++ achieves a significantly higher level of segmentation performance over U-Net across all the biomedical applications under study. 
}}
\label{tab:appendix_main_results}
{\jlrev
\begin{tabular}{p{0.19\linewidth}P{0.14\linewidth}P{0.14\linewidth}P{0.14\linewidth}P{0.14\linewidth}}
    \hline
    EM & Sensitivity & Specificity & F1 score & F2 score \\
    \hline
    U-Net & 91.21{\tiny $\pm$2.18} & 83.55{\tiny $\pm$1.62} & 87.21{\tiny $\pm$1.88} & 89.56{\tiny $\pm$2.06} \\
    UNet++ & 92.87{\tiny $\pm$2.08} & 84.94{\tiny $\pm$1.55} & 88.73{\tiny $\pm$1.79} & 91.17{\tiny $\pm$1.96} \\
    $p$-value & 0.018 & 0.008 & 0.013 & 0.016 \\
    \hline
    \hline
    Cell & Sensitivity & Specificity & F1 score & F2 score \\
    \hline
    U-Net & 94.04{\tiny $\pm$2.36} & 96.10{\tiny $\pm$0.75} & 81.25{\tiny $\pm$2.62} & 88.47{\tiny $\pm$2.49} \\
    UNet++ & 95.88{\tiny $\pm$2.59} & 96.76{\tiny $\pm$0.65} & 84.34{\tiny $\pm$2.52} & 90.90{\tiny $\pm$2.57} \\
    $p$-value & 0.025 & 0.005 & 5.00$e$-4 & 0.004 \\
    \hline
    \hline
    Nuclei & Sensitivity & Specificity & F1 score & F2 score \\
    \hline
    U-Net & 93.57{\tiny $\pm$4.30} & 93.94{\tiny $\pm$0.87} & 83.64{\tiny $\pm$2.97} & 89.33{\tiny $\pm$3.71} \\
    UNet++ & 97.28{\tiny $\pm$4.85} & 96.30{\tiny $\pm$0.94} & 90.14{\tiny $\pm$3.82} & 94.29{\tiny $\pm$4.41} \\
    $p$-value & 0.015 & 5.35$e$-10 & 6.75$e$-7 & 4.47$e$-4 \\
    \hline
    \hline
    Brain Tumor & Sensitivity & Specificity & F1 score & F2 score \\
    \hline
    U-Net & 94.00{\tiny $\pm$1.15} & 97.52{\tiny $\pm$0.78} & 88.42{\tiny $\pm$2.61} & 91.68{\tiny $\pm$1.77} \\
    UNet++ & 95.81{\tiny $\pm$1.25} & 98.01{\tiny $\pm$0.67} & 90.83{\tiny $\pm$2.46} & 93.75{\tiny $\pm$1.77} \\
    $p$-value & 2.90$e$-5 & 0.042 & 0.005 & 7.03$e$-3 \\
    \hline
    \hline
    Liver & Sensitivity & Specificity & F1 score & F2 score \\
    \hline
    U-Net & 91.22{\tiny $\pm$2.02} & 98.48{\tiny $\pm$0.43} & 86.19{\tiny $\pm$2.84} & 89.14{\tiny $\pm$2.37} \\
    UNet++ & 93.15{\tiny $\pm$1.88} & 98.74{\tiny $\pm$0.36} & 88.54{\tiny $\pm$2.57} & 91.25{\tiny $\pm$2.18} \\
    $p$-value & 0.003 & 0.046 & 0.010 & 0.006 \\
    \hline
    \hline
    Lung Nodule & Sensitivity & Specificity & F1 score & F2 score \\
    \hline
    U-Net & 94.95{\tiny $\pm$1.31} & 97.27{\tiny $\pm$0.47} & 83.98{\tiny $\pm$1.94} & 90.24{\tiny $\pm$1.60} \\
    UNet++ & 95.83{\tiny $\pm$0.86} & 97.81{\tiny $\pm$0.40} & 86.78{\tiny $\pm$1.66} & 91.99{\tiny $\pm$1.22} \\
    $p$-value & 0.018 & 3.25$e$-3 & 1.92$e$-5 & 4.27$e$-3 \\
    \hline
    \end{tabular}
}
\end{threeparttable}
\end{center}
\end{table}

}

\newpage
{\jlrev
\section{Learning curves}
\label{sec:learning_curves}

}

%##############################################################################################
\begin{figure}[h]
\begin{center}
\includegraphics[width=1.0\linewidth]{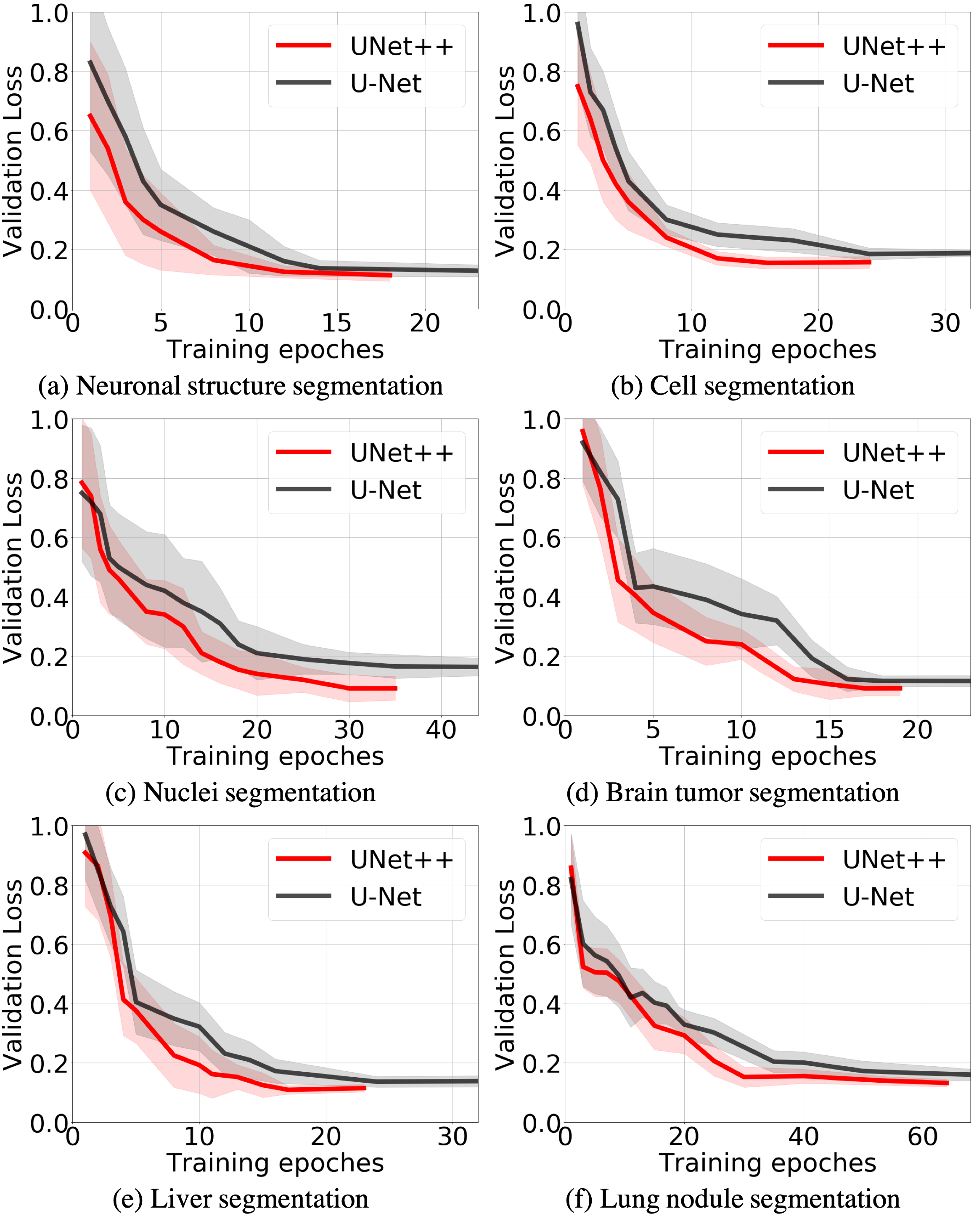}
\end{center}
\caption{{\jlrev UNet++ enables a better optimization than U-Net evidenced by the learning curves for the tasks of neuronal structure, cell, nuclei, brain tumor, liver, and lung nodule segmentation. We have plotted the validation losses averaged by 20 trials for each application. As seen, UNet++ with deep supervision accelerates the convergence speed and yields the lower validation loss due to the new design of the intermediate layers and dense skip connections.}}
\label{fig:learning_curves}
\end{figure}
%##############################################################################################

\end{document}